\documentclass[aps,pre,showpacs,superscriptaddress]{revtex4}

\usepackage{amsfonts}                   
\usepackage{amssymb}
\usepackage{amsbsy}

\newcommand{\BEQ}{\begin{equation}}     
\newcommand{\BEA}{\begin{eqnarray}}
\newcommand{\EEQ}{\end{equation}}       
\newcommand{\EEA}{\end{eqnarray}}
\newcommand{\eps}{\varepsilon}          
\newcommand{\sig}{\sigma}               
\newcommand{\D}{{\rm d}}                
\newcommand{\II}{{\rm i}}               
\newcommand{\wit}[1]{\widetilde{#1}}    

\renewcommand{\vec}[1]{{\bf{#1}}}       

\newcommand{\zeile}[1]{\vskip #1 \baselineskip} 
\newcommand{\appsection}[1]{\setcounter{equation}{0} \section*{Appendix #1}
\renewcommand{\theequation}{A\arabic{equation}}
              \renewcommand{\thesection}{A} }

\begin{document}

\input epsf.sty

\widetext

\title{Scaling of the linear response in simple ageing systems without disorder}

\author{Malte Henkel}
\affiliation{Laboratoire de Physique des Mat\'eriaux,$^{*}$
Universit\'e Henri Poincar\'e Nancy I, B.P. 239,\\
F -- 54506 Vand{\oe}uvre l\`es Nancy Cedex, France}

\author{Matthias Paessens}
\affiliation{Institut f\"ur Festk\"orperforschung (Theorie II), 
Forschungszentrum J\"ulich, D -- 52425 J\"ulich, Germany}

\author{Michel Pleimling}
\affiliation{Institut f\"ur Theoretische Physik I, 
Universit\"at Erlangen-N\"urnberg, D -- 91058 Erlangen, Germany}

\begin{abstract}
The time-dependent scaling of the thermoremanent and zero-field-cooled
susceptiblities in ferromagnetic spin systems undergoing ageing after a quench
to a temperature at or below criticality is studied. A recent debate 
on their interpretation is resolved by showing that for systems with a
short-ranged equilibrium spin-spin correlator and above their roughening 
temperature, the field-cooled susceptibility
$\chi_{\rm FC}(t)-\chi_0\sim t^{-A}$ where $\chi_0$ is related to the 
equilibrium magnetization and the exponent $A$ is related to the 
time-dependent scaling of the interface width between ordered domains. The
same effect also dominates the scaling of the zero-field-cooled susceptibility
$\chi_{\rm ZFC}(t,s)$, but does not enter into the thermoremanent susceptibility
$\rho_{\rm TRM}(t,s)$. However, there may be large finite-time corrections
to the scaling of $\rho_{\rm TRM}(t,s)$ which are explicitly derived and 
may be needed in order to extract reliable ageing exponents. Consistency with
the predictions of local scale invariance is confirmed in the Glauber-Ising
and spherical models.

\zeile{1}
\end{abstract}
\maketitle

\section{Introduction}

The comprehension of the physics of ageing phenomena is a topic of much
current interest. While ageing was originally studied in glassy systems, there
are many conceptual problems which are conveniently studied first in simple
ferromagnetic systems. In this paper, we shall study a ferromagnet with
a critical temperature $T_c>0$, initially prepared in a fully disordered 
(infinite-temperature) state which is quenched at time $t=0$ to a temperature 
$T\leq T_c$. We shall consider throughout a dynamics with a non-conserved order
parameter. Physically, the ageing process proceeds via the growth of correlated
domains of size $L(t)\sim t^{1/z}$ and the slow motion of the
domain boundaries drives in turn the slow temporal evolution of macroscopic
observables, see \cite{Bou00,God02,Cug02} for recent reviews. It has turned 
out that ageing phenomena are more fully revealed through the study of 
two-time quantities such as the two-time correlator $C(t,s)$ and the two-time 
linear response function $R(t,s)$ defined by
\BEQ
C(t,s) = \left\langle \phi(t) \phi(s)\right\rangle \;\; , \;\;
R(t,s) = \left.\frac{\delta \langle\phi(t)\rangle}{\delta h(s)}\right|_{h=0}
\EEQ
where $\phi(t)$ is the time-dependent order parameter, $h(s)$ is the magnetic 
field conjugate to $\phi$, $t$ is referred to as observation time and $s$ will 
be called the waiting time. Causality implies that $R(t,s)=0$ for $t<s$. 

Ageing systems may display dynamical scaling in the long-time limit 
\cite{Bou00,God02,Cug02}. Specifically,
consider the two-time functions in the ageing regime $t\gg t_{\rm micro}$,
$s\gg t_{\rm micro}$ and $\tau =t-s\gg t_{\rm micro}$, where $t_{\rm micro}$ 
is some microscopic time. Then one has the scaling behaviour
\BEQ \label{1:gl:CRskal}
C(t,s) \sim s^{-b} f_{C}(t/s) \;\; , \;\;
R(t,s) \sim s^{-1-a} f_{R}(t/s)
\EEQ
where the scaling functions $f_{C,R}(x)$ have the following asymptotic 
behaviour for $x\to\infty$
\BEQ
f_{C}(x) \sim x^{-\lambda_C/z} \;\; , \;\;
f_{R}(x)\sim x^{-\lambda_R/z}
\EEQ
and $\lambda_C$ and $\lambda_R$ are the autocorrelation \cite{Fis88,Hus89} and 
autoresponse \cite{Pic02} exponents, respectively. In general, the exponents 
$\lambda_{C,R}$ and $z$ will take different values for $T<T_c$ and for $T=T_c$. 
In particular, $z=2$ for $T<T_c$ and a non-conserved order parameter.

The values of the exponents $a$ and $b$ are collected in table~\ref{tab1} and
depend on the equilibrium spin-spin correlator $C_{\rm eq}$ as follows 
\cite{Hen02a}. If $C_{\rm eq}(\vec{r})\sim e^{-|\vec{r}|/\xi}$ with a finite 
$\xi$, one says that the system is of {\em class S}, while if
$C_{\rm eq}(\vec{r})\sim |\vec{r}|^{-(d-2+\eta)}$, the system is said to
be of {\em class L}, where $\eta$ is a standard equilibrium critical exponent.
We point out that for systems of class S, 
the result $a=1/z$ follows from the
well-accepted intuitive picture that ageing effects come from the slow motion of
the domains walls which separate the well-ordered domains in systems undergoing
coarsening \cite{Ber99,Bou00,Cug02,Hen02a}. A different value for $a$ would 
invalidate this physical picture. 

Recently, it has been proposed that ageing systems might possess a larger
dynamically generated space-time symmetry than mere dynamical scaling 
\cite{Hen02}. Indeed, for any given value of $z$, infinitesimal local scale 
transformations with a  space-time-dependent rescaling factor 
$1+\eps(t,\vec{r})$ can be constructed. 
In particular, the following explicit expression for the response function
is obtained from the condition that $R(t,s)$ transforms covariantly under 
the action of local scale transformations \cite{Hen02,Hen01}
\BEQ \label{1:gl:lsi}
R(t,s) = r_0' \left(\frac{t}{s}\right)^{1+a-\lambda_R/z} 
(t-s)^{-1-a}
\EEQ
and where $r_0'$ is a normalization constant. This prediction has been 
confirmed in several models, notably the kinetic Ising model with
Glauber dynamics \cite{Hen01,Hen03b} and several variants of the
exactly solvable spherical model \cite{Hen01,God00b,Can01}, as is reviewed
in \cite{Hen02,Hen03c}. Furthermore, predictions exist for the spatio-temporal
response. A particular simple result is found for the case $z=2$
\cite{Hen02,Hen03}
\BEQ \label{1:gl:Rrz}
R(t,s;\vec{r}) = \left.\frac{\delta \langle\phi_{\vec{r}}(t)\rangle}
{\delta h_{\vec{0}}(s)}\right|_{h=0} = R(t,s) \exp\left[-\frac{{\cal M}}{2}
\frac{\vec{r}^2}{t-s}\right]
\EEQ
($\cal M$ is a non-universal constant) and has been confirmed in the $2D$ and 
$3D$ Glauber-Ising model quenched to $T<T_c$ \cite{Hen03c} (and also in the
spherical model for $d>2$). Since the derivation of (\ref{1:gl:Rrz}) depends on
the Galilei invariance of the coarsening process, its verification is strong 
evidence in favour of Galilei invariance as a dynamically generated space-time 
symmetry of phase-ordering.

One would obviously like to be able to test statements of such a general
nature as widely as possible, but 
since there are so few exactly solvable and non-trivial model of ageing,
quantities such as $R(t,s)$ must in general be obtained numerically. This paper
addresses an important conceptual point on the interpretation of the scaling
behaviour of several commonly used observables related to $R(t,s)$ and we 
shall try to clear up a debate which has arisen in recent years.  

Numerical information on $R(t,s)$ can be obtained using a by now standard 
method devised by Barrat \cite{Bar98}. One perturbs the system by a {\em random}
magnetic field $h_{\vec{i}}$ with zero mean $\overline{h_{\vec{i}}}=0$. We 
shall use a binary field $h_i = \pm h$ below in our Glauber-Ising model 
simulation, but a gaussian random field is also possible. 
Instead of measuring $R(t,s)$ directly, two common procedures run as follows.
Either one quenches the system and turns on the magnetic field after the
waiting time $s$ has elapsed and then works with the zero-field-cooled 
susceptibility $\chi_{\rm ZFC}(t,s)$. Alternatively, one may also keep 
the random field until the waiting time $s$ when it is turned off and then
has the thermoremanent susceptibility $\rho_{\rm TRM}(t,s)$. These 
are related to $R(t,s)$ as follows
\BEA 
\chi_{\rm ZFC}(t,s) &=& \chi(t,s) = 
\int_{s}^{t}\!\D u\, R(t,u) \stackrel{?}{=} s^{-a} f_{\chi}(t/s) 
\label{1:gl:chiskal} \\
\rho_{\rm TRM}(t,s) &=& \rho(t,s) = 
\int_{0}^{s}\!\D u\, R(t,u) \stackrel{?}{=} s^{-a} f_{M}(t/s)
\label{1:gl:rhoskal}
\EEA
Here, we straightforwardly used the scaling forms (\ref{1:gl:CRskal})
{\em without paying attention to the conditions of validity} of these and in
particular {\em did not pay any attention as to whether $t-s\gg t_{\rm micro}$ 
holds true or not}. 
As we shall show, however, careful consideration of these conditions is crucial 
in order to obtain valid scaling forms for the integrated responses 
$\chi_{\rm ZFC}(t,s)$ and  $\rho_{\rm TRM}(t,s)$. 

On a discrete lattice $\Lambda\subset \mathbb{Z}^d$, the integrated responses 
(\ref{1:gl:chiskal},\ref{1:gl:rhoskal}) are obtained by measuring the 
time-dependent magnetization \cite{Bar98}
\BEQ \label{1:gl:Barrat}
M = \frac{1}{|\Lambda|\, h}
\overline{\left\langle\sum_{\vec{i}\in\Lambda} 
\phi_{\vec{i}}(t) h_{\vec{i}}\right\rangle}
\EEQ
where $|\Lambda|$ is the number of sites of the lattice $\Lambda$. Depending 
on whether one works in the zero-field-cooled or the thermoremanent protocol, 
one obtains $\chi_{\rm ZFC}=M_{\rm ZFC}/h$ or $\rho_{\rm TRM}=M_{\rm TRM}/h$, 
respectively.

In recent years, Corberi, Lippiello 
and Zannetti (CLZ) \cite{Cor01,Cor03} have studied in great detail ageing in 
simple ferromagnets such as the Glauber-Ising model. Most notably, and based 
on MC data on $\chi_{\rm ZFC}$ in between one and four dimensions, 
they have argued that for the phase-ordering kinetics (i.e. $T<T_c$) in the 
Glauber-Ising model, the exponent $a$ takes a value different from
the generally accepted value $a=1/z=1/2$. Their studies consider the two-time
autocorrelation $C(t,s)$ and the dynamic susceptibilities $\chi(t,s),\rho(t,s)$ 
of the O($n$) vector model. They separate the autocorrelator
into a `stationary' and an `ageing' part 
$C(t,s) = C^{({\rm st})}(t-s)+C^{({\rm age})}(t,s)$ and similarly 
$\chi(t,s) = \chi^{({\rm st})}(t-s)+\chi^{({\rm age})}(t,s)$, where 
$\chi^{({\rm st})}(t,s)$ is {\em defined} such as to satisfy the 
fluctuation-dissipation theorem with $C^{({\rm st})}(t,s)$ (and similarly for
$\rho(t,s)$). Furthermore, CLZ admit dynamical scaling for the `ageing' part 
only and further introduce two distinct exponents $a_{\chi}$ and $a_{\rho}$ 
according to
\BEQ
\chi^{({\rm age})}(t,s) \sim s^{-a_{\chi}} f_{\chi}(t/s) \;\; , \;\;
\rho^{({\rm age})}(t,s) \sim s^{-a_{\rho}} f_{\rho}(t/s) 
\EEQ
If $\hat{a}_n$ is the value of $a_{\chi}$ in the O($n$)-model, CLZ propose 
\BEQ \label{1:gl:aHut}
\hat{a}_1 = \left\{\begin{array}{ll} 
(d-1)/4 & \mbox{\rm ~~;~~ $d<3$} \\
1/2     & \mbox{\rm ~~;~~ $d>3$}
\end{array} \right.
\;\; ; \;\;
\hat{a}_{\infty} = \left\{\begin{array}{ll} 
(d-2)/2 & \mbox{\rm ~~;~~ $d<4$} \\
1       & \mbox{\rm ~~;~~ $d>4$}
\end{array} \right.
\EEQ
for the Glauber-Ising ($n=1$) and spherical model ($n=\infty$), respectively
(additional logarithmic factors may occur at 
$d=3, n=1$ and $d=4, n=\infty$, respectively). 
In order to account for these results, CLZ invoke a dangerous irrelevant 
variable and claim that $d=3$ were a critical dimension of coarsening in the
Glauber-Ising model. 
On the other hand, from a log-log plot of $\rho^{({\rm age})}(t,s)$ 
against $s$ in the Glauber-Ising model 
{\it ``\ldots no statement on $a_{\rho}$ can be made \ldots''} \cite{Cor03},
but roughly $a_{\rho}\approx 0.55-0.60$ in $2D$ and 
$a_{\rho}\approx 0.65-0.70$ in $3D$, see \cite[Fig. 6]{Cor03}. 
These conclusions are supplemented by a discussion of the 
necessary condition $s\gg t_{\rm micro}$ needed for the validity of the
scaling (\ref{1:gl:CRskal}) and about the microscopic times $t_{\rm micro}$
which may enter there \cite{Cor03}. However, the crucial r\^ole of the 
additional condition $t-s\gg t_{\rm micro}$ which must also be satisfied
for eq.~(\ref{1:gl:CRskal}) to hold is not addressed by CLZ.

Already from a pure phemomenological point of view and given the scaling
relations (\ref{1:gl:CRskal}) and the definitions of $\chi(t,s)$ and 
$\rho(t,s)$, the assertion of CLZ of having {\em distinct} exponents 
$a_{\chi}\ne a_{\rho}$ cannot be maintained. 
Rather, it is necessarry to go beyond a pure scaling analysis in order to 
understand how the ageing behaviour can be correctly extracted from 
integrated response functions. 

Our argument runs along the following lines.
\begin{enumerate}
\item 
We consider the field-cooled susceptibility
\BEQ
\chi_{\rm FC}(t) = \chi_{\rm ZFC}(t,s) + \rho_{\rm TRM}(t,s) = 
\int_{0}^{t} \!\D u\, R(t,u)
\EEQ
For ferromagnets of class S, one has $\chi_{\rm FC}(t)-\chi_0\sim t^{-A}$, 
as we shall show in section~\ref{sect2}. 
Here $T \, \chi_0=1-m_{\rm eq}^2$ is given
by the mean equilibrium magnetization. The exponent $A$ is a new 
exponent without a direct relationship to ageing, rather it is related to the 
roughness of the interface between ordered domains. We stress that since the 
power-law behaviour of $\chi_{\rm FC}(t)$ is {\em independent} of the waiting 
time $s$ it has no relation with a possible ageing behaviour.
In particular, we find $A=1/4$ for the $2D$ Glauber-Ising model with $T<T_c$. 
Furthermore, for a fixed scaling variable $x=t/s$ one has, with the scaling 
function $g(x) \sim x^{-A}$
\BEQ \label{1:gl:chiZFCskal}
\chi_{\rm ZFC}(t,s) = \chi_{\rm FC}(t) - \rho_{\rm TRM}(t,s) \sim 
\chi_0 + s^{-A} g(x) + {\rm O}(s^{-a})
\EEQ
Here we anticipate an important result of section~\ref{sect4}, namely that
$\rho_{\rm TRM}(t,s)\sim s^{-a}$, see also eq.~(\ref{1:gl:rhoTRMskal}) below. 
Since for rough interfaces one has $A-a=A-1/z<0$, 
it follows that (\ref{1:gl:chiskal}) cannot be used. Furthermore, the
splitting of $\chi(t,s)$ into a `stationary' and an `ageing' part advocated by
CLZ is in contradiction with (\ref{1:gl:chiZFCskal}). Rather, the leading 
scaling of $\chi_{\rm FC}$ with the waiting time
$s$ is unrelated to the ageing behaviour of the model as described by
(\ref{1:gl:CRskal}). Indeed, the terms 
describing ageing only occur as subleading terms in $\chi_{\rm ZFC}(t,s)$ and 
are therefore difficult to extract. In particular, the
exponent $A$ cannot be identified with the ageing exponent $a$. 
An example is provided by the $2D$ Glauber-Ising model, 
where $a=1/z=1/2$ but $A=1/4$. 
\item 
For ferromagnets of class L, we shall show that $A=0$. We shall explicitly
test this in section~\ref{sect3} for the non-equilibrium critical dynamics 
(i.e. $T=T_c$) in the $2D$ Glauber-Ising model and in the 
kinetic mean spherical model for any $T\leq T_c$. 
\item 
Eq.~(\ref{1:gl:chiZFCskal}) shows that ageing effects merely provide a 
finite-time correction to the scaling of $\chi_{\rm ZFC}$. On the other hand, 
ageing terms are leading in the thermoremanent magnetization which scales 
more precisely as \cite{Hen02a}
\BEQ \label{1:gl:rhoTRMskal}
\rho_{\rm TRM}(t,s) = s^{-a} f_{M}(t/s) + s^{-\lambda_R/z} g_M(t/s)
\EEQ 
where the scaling functions $f_M(x)$ and $g_M(x)$ are related to the 
response function $R(t,s)$ and can be found explicitly, see section~\ref{sect4}.
For example, for a system in class S with an uncorrelated 
initial state, one has 
$\lambda_C=\lambda_R\geq d/2$ \cite{Yeu96} and since $a=1/z$, these two terms
may be of almost the same order and a simple log-log plot may not be
sufficient to yield a precise value of $a$ for times accessible in present
simulations. Indeed, this situation occurs in the $2D$ and $3D$  
Glauber-Ising model. However, subtracting the leading correction
according to eq.~(\ref{1:gl:rhoTRMskal}) allows to reliably determine $a$ and 
the scaling function $f_M(x)$ \cite{Hen02a,Hen03b}. We shall describe this 
in section~\ref{sect4}.
\end{enumerate}
In an appendix possible implications for the scaling of $R(t,t-\eps)$ for 
$\eps\to 0$ are discussed.

\section{Scaling of $\chi_{\rm FC}$ for short-ranged equilibrium
correlators} \label{sect2}

We begin by discussing the time-dependent scaling for the field-cooled
susceptibility $\chi_{\rm FC}(t)$ for the Glauber-Ising model. We consider 
the Ising model on a hypercubic lattice with periodic boundary conditions and
the equilibrium Hamiltonian 
${\cal H}=-\sum_{(\vec{i},\vec{j})} \sig_{\vec{i}}\sig_{\vec{j}}$ where the
sum is over nearest neighbours only. We use heat-bath dynamics defined through 
the stochastic rule
\BEQ
\sig_{\vec{i}}(t+1) = \pm 1 \;\; \mbox{\rm with probability
$\frac{1}{2}\left[ 1\pm\tanh\left( H_{\vec{i}}(t)/T\right)\right]$}
\EEQ
with the local field $H_{\vec{i}}(t)=\sum_{\vec{y}(\vec{i})}\sig_{\vec{y}}(t)$
and where $\vec{y}(\vec{i})$ runs over the nearest neighbours of the 
sites $\vec{i}$. Initially, the system is prepared in an infinite-temperature
state. Thermoremanent and zero-field-cooled susceptibilities
$\rho_{\rm TRM}(t,s)$ and $\chi_{\rm ZFC}(t,s)$ can now be measured
by perturbing the model by a binary random field $h_{\vec{i}}=\pm h$ with
zero mean $\overline{h_{\vec{i}}}=0$ and using eq.~(\ref{1:gl:Barrat}). From
these two independent measurements, we obtain the field-cooled
susceptibility
\BEQ
\chi_{\rm FC}(t) = \rho_{\rm TRM}(t,s) + \chi_{\rm ZFC}(t,s)
\EEQ
In figure~\ref{Abb1}, we show $\chi_{\rm FC}(t)$ so obtained for several
values of the waiting time $s$ in the case of the two-dimensional Ising model, 
at the fixed temperature $T=1.5<T_c$. 
Clearly, $\chi_{\rm FC}$ is independent of $s$, which means that we are well
inside the linear-response regime. Furthermore, within the times considered, 
our data are consistent with the scaling behaviour
\BEQ \label{2:gl:chiFCskal}
\chi_{\rm FC}(t) \sim t^{-A} \;\; , \;\; A\approx 0.25
\EEQ
Since this dynamical scaling is completely independent of the waiting time $s$,
it is unrelated to any ageing behaviour which might occur in this model. 

We now give a heuristic argument in order to understand where the scaling form
(\ref{2:gl:chiFCskal}) and the value of $A$ come from. Consider a simple 
ferromagnet in $d>1$ dimensions 
which is quenched at time $t=0$ from an infinite-temperature initial
state to a final temperature $T<T_c$. The dynamics is assumed to be purely 
relaxational, i.e. without any conservation law. Microscopically, it is 
well-known that the configurations of the system consist of fully ordered 
domains of spins, of a typical size $L(t)\sim t^{1/z}$ with $z=2$.  
We now perturb with a random 
field of zero mean $\overline{h_{\vec{i}}}=0$ and wish to obtain the
susceptibility $\chi_{\rm FC}=M_{\rm FC}/h$ from eq.~(\ref{1:gl:Barrat}). 
First consider the case when $T=0$. Then, because the spins deep inside 
the cluster are ordered, the only 
non-vanishing contribution to $\chi_{\rm FC}$ comes from the spins from near 
the interfaces between the ordered clusters. We denote the interface density
by $\rho_{\rm I}(t)$ and have $\rho_{\rm I}(t)\sim L(t)^{-1}$, 
see \cite{Ber99,Bou00,Cug02}. If $w(t)$ is the interface width, we have
\BEQ \label{2:gl:grenz}
\chi_{\rm FC}(t) = \frac{1}{|\Lambda|\, h^2}\overline{\left\langle 
\sum_{\vec{i}\in\Lambda} \sig_{\vec{i}}(t) h_{\vec{i}}\right\rangle}
= \frac{1}{|\Lambda|\, h^2}\overline{\left\langle 
\sum_{\,\vec{i}\in\,{\rm interfaces}}
\sigma_{\vec{i}}(t) h_{\vec{i}}\right\rangle}
\sim L(t)^{-1} w(t)
\EEQ
For a finite temperature $T>0$, the order deep inside the clusters is not 
perfect and there remains a residual contribution to the susceptiblity. 
We then have, for large times
\BEQ
\chi_{\rm FC}(t) \simeq \chi_{0} + L(t)^{-1} w(t) \;\; , \;\;
T \chi_0 = 1 - m_{\rm eq}^2
\EEQ
where $m_{\rm eq}$ is the equilibrium magnetization (for the $2D$ Ising model,
$m_{\rm eq}=(1-\sinh(2/T)^{-4})^{1/8}$, see e.g. \cite{Bax82}). 
The dynamics of a $(d-1)$-dimensional interface in a 
$d$-dimensional system ($d \geq 2$)
can be described by the dynamics of a height model \cite{Abr89} of continuous 
height variables $v_{\vec{j}}\in\mathbb{R}$ and the equilibrium Hamiltonian
${\cal H}[v] = -\frac{\tau}{2}\sum_{(\vec{j},\vec{j}')} 
\left( v_{\vec{j}}-v_{\vec{j}'}\right)^2$ with nearest-neighbour interactions 
and where $\tau$ is the effective interfacial tension \cite{Abr89}. 
In adopting this description, we tacitly assume that the system is above
its `roughening temperature' $T_R$, see e.g. \cite{Vil86}, such that the
fluctuations of the interface are unbounded (rough interface) for $T>T_R$ and 
with bounded fluctuations for $T<T_R$ (smooth interface). This condition
is always satisfied for $d=2$, since then $T_R=0$ and indeed the
description adopted here can be derived rigorously \cite{Abr81}. 
On the other hand, for $d=3$ one has $T_R\approx 0.5\, T_c$ and finally, 
$T_R=\infty$ for $d\geq 4$. If the dynamics of the model is described by a 
Langevin equation, the squared interface width was shown by 
Abraham and Upton to scale for 
large times as $w(t)^2 = \langle v_{\vec{0}}(t)^2\rangle \sim t^{1/2}$ in $2D$ 
and $w(t)^2 \sim \ln t$ in $3D$ \cite{Abr89}. Inserting this
into eq.~(\ref{2:gl:grenz}), we find the
following leading dynamical scaling for $\chi_{\rm FC}$ in the 
Glauber-Ising model as $t\to\infty$
\BEQ \label{2:gl:A}
\chi_{\rm FC}(t) -\chi_0 \sim \left\{
\begin{array}{ll} t^{-1/4}               & \mbox{\rm ~~;~~ if $d=2$} \\
                  t^{-1/2} (\ln t)^{1/2} & \mbox{\rm ~~;~~ if $d=3$}
\end{array} \right. 
\EEQ
and provided that $T>T_R$. 
For dimensions $d\geq 4$, and generically if $T<T_R$, 
one expects a flat interface with 
$w(t)\sim {\rm const.}$ and consequently 
$\chi_{\rm FC}(t) - \chi_0 \sim t^{-1/2}$. Reconsidering figure~\ref{Abb1}, 
it is easy to check that the final approch of $\chi_{\rm FC}$ towards
$\chi_0$ only occurs at times much beyond those accessed by our simulation.  

Combining eqs.~(\ref{1:gl:chiZFCskal}) and (\ref{2:gl:A}), 
we have therefore reproduced the findings of CLZ in the Glauber-Ising model
and have also made clear the physical origin of these results. Besides,
a definite prediction for the logarithmic factor in $d=3$ was obtained. 
In summary, we have for the Glauber-Ising model $A=1/4$ in $2D$ and
$A=1/2$ for all $d\geq 3$, up to a known logarithmic correction in $3D$.

\section{Scaling of $\chi_{\rm FC}$ for long-ranged equilibrium
correlators} \label{sect3}

We now ask whether the heuristic discussion of the scaling of $\chi_{\rm FC}$ 
presented in the previous section can be taken over for systems of class L. 
Indeed, the main physical difference with respect to systems of class S is that
although correlated clusters of size $L(t)\sim t^{1/z}$ 
form, fluctuations do persist in the interior of these clusters on all length
scales up to $L(t)$. This means that one should consider an `interface  width' 
scaling as $w(t)\sim L(t)$. This in turn leads to 
$\chi_{\rm FC}(t)\sim {\rm const.}$ and $A=0$ (on the other hand, since the
clusters should have no `inside', we do not expect a term $\chi_0$ to occur). 

We now test this heuristic idea in the exactly solvable mean spherical model 
and shall also present evidence from the $2D$ Glauber-Ising model quenched 
onto criticality.  

{\bf 1.} First, we consider the mean kinetic spherical 
model, see e.g. \cite{Cug95,God00b,Can01,Pic02,Pae03}. 
To each site $\vec{x}$ of a hypercubic lattice 
$\Lambda\subset\mathbb{Z}^d$ one attaches a continuous spin variable
$S_{\vec{x}}(t)\in\mathbb{R}$, subject to the mean spherical constraint
\BEQ \label{3:gl:mszw}
\sum_{\vec{x}\in\Lambda} \left\langle S_{\vec{x}}^2\right\rangle =1 
\EEQ
By analogy with the Glauber-Ising model, we also add a random magnetic field
$h_{\vec{x}}$ and the equations of motion read
\BEQ \label{3:gl:equmo}
\frac{\D}{\D t} S_{\vec{x}}(t) = \sum_{\vec{y}(\vec{x})} S_{\vec{y}}(t) 
-(2d-\mathfrak{z}(t))S_{\vec{x}}(t) + h_{\vec{x}} +\eta_{\vec{x}}(t)
\EEQ
where the noise and the random field have zero average
$\langle \eta_{\vec{x}}\rangle = \overline{h_{\vec{x}}} = 0$ and the
correlators
\BEQ \label{3:gl:varianzen}
\left\langle \eta_{\vec{x}}(t)\eta_{\vec{y}}(t')\right\rangle 
= 2 T\, \delta_{\vec{x},\vec{y}}\, \delta(t-t') \;\; , \;\;
\overline{h_{\vec{x}}h_{\vec{y}}} 
= 2 \Gamma\, \delta_{\vec{x},\vec{y}} 
\EEQ
where the temperature $T$ and the width $\Gamma$ are constants and
$\mathfrak{z}(t)$ is a Lagrange multiplier to be determined below. In addition,
the noise and the field are assumed independent, i.e.\ 
$\overline{\langle \eta_{\vec{x}}(t) h_{\vec{y}}\rangle}=0$, and in addition 
the initial state is uncorrelated in the sense that 
$\overline{\langle S_{\vec{x}}(0)h_{\vec{y}}\rangle}=0$. 
Here and in the following the average $\langle \mathfrak{X}\rangle$ is always 
taken over the initial conditions and the noise, while the average 
$\overline{\,\mathfrak{X}\,}$ is over the random field. The solution of this 
model follows standard lines \cite{Pae03}. 
Taking Fourier transforms, the solution of eq.~(\ref{3:gl:equmo}) is
\BEQ \label{3:gl:ssol}
\wit{S}(\vec{q},t) = \frac{\exp(-\omega(\vec{q})t)}{\sqrt{g(t)}} 
\left[ \wit{S}(\vec{q},0) + \int_0^t \!\D t'\,
e^{\omega(\vec{q})t'}\sqrt{g(t')}\left( \wit{h}(\vec{q})+\wit{\eta}(\vec{q},t')
\right) \right]
\EEQ
with the dispersion relation $\omega(\vec{q})=2\sum_{i=1}^d(1-\cos q_i)$ and 
$g(t)=\exp(2\int_0^t\!\D t'\, \mathfrak{z}(t'))$. 
The spin-spin 
correlator $\wit{C}(\vec{q},\vec{q}';t,s)=
\overline{\left\langle \wit{S}(\vec{q},t)\wit{S}(\vec{q}',s)\right\rangle}$
is readily found and we have in direct space the autocorrelator
\BEA
\lefteqn{ 
C(t,s) = C_{\vec{x},\vec{x}}(t,s) = 
\frac{1}{(2\pi)^{2d}} \int_{{\cal B}^2} \!\D\vec{q}\D\vec{q}'\:
e^{\II (\vec{q}+\vec{q}')\cdot \vec{x}}\, \wit{C}(\vec{q},\vec{q}';t,s)
}
\nonumber \\
&=& \frac{1}{\sqrt{g(t)g(s)\,}} 
\left[ A\left(\frac{t+s}{2}\right)+2T \int_0^s\!\D u\, f\left(\frac{t+s}{2}-u
\right) g(u) + 2\Gamma \int_{0}^t \!\D t' \int_0^s \!\D s' \: 
f\left(\frac{t+s-t'-s'}{2}\right) \sqrt{ g(t') g(s')\,}\right] \qquad
\label{3:gl:Cts}
\EEA
where $\cal B$ is the Brillouin zone and with the definitions
\BEA
A(t) &=& (2\pi)^{-d} \int_{\cal B} \!\D \vec{q}\: e^{-2\omega(\vec{q})t}
\wit{C}(\vec{q},0) 
\\
f(t) &=& (2\pi)^{-d} \int_{\cal B} \!\D \vec{q}\: e^{-2\omega(\vec{q})t}
= \left( e^{-4t} I_0(4t)\right)^d
\EEA
and where $I_0$ is a modified Bessel function. For infinite-temperature initial
conditions $A(t)=f(t)$. The mean spherical constraint (\ref{3:gl:mszw}) gives
$C(t,t)=1$ and this leads to the following generalized Volterra integral 
equation
\BEQ \label{3:gl:g}
g(t) = A(t) + 2T \int_{0}^t \!\D u\, f(t-u) g(u) 
+2\Gamma \int_0^t \!\D u' \int_0^t \D u''\: f\left(t-\frac{u'+u''}{2}\right)
\sqrt{g(u')g(u'')\,} 
\EEQ
which determines $g(t)$. Finally, the response function is given by the
usual equation $R(t,s) = f((t-s)/2)\sqrt{g(s)/g(t)}$. At zero temperature $T=0$,
eqs.~(\ref{3:gl:Cts},\ref{3:gl:g}) are identical to those found for
the spherical spin-glass \cite{Cug95,Ber01}. 

The field-cooled susceptibility is given by
\BEQ \label{3:gl:chifc}
\chi_{\rm FC}(t) = \int_0^t \!\D u\, R(t,u) = \int_0^t \!\D u\,
f\left(\frac{t-u}{2}\right) \sqrt{\frac{g(u)}{g(t)}\,}
\EEQ
At this point, it is instructive to re-derive the equivalence
between the definition (\ref{3:gl:chifc}) and eq.~(\ref{1:gl:Barrat}),
originally proposed \cite{Bar98} for the Glauber-Ising model, in the context 
of the mean kinetic spherical model. Indeed, we have ($|\Lambda|$ denotes the
number of sites of the lattice $\Lambda$) 
\BEA
\frac{1}{|\Lambda|} 
\overline{\left\langle \sum_{\vec{x}\in\Lambda} S_{\vec{x}}(t) h_{\vec{x}}
\right\rangle} &=& \frac{(2\pi)^{-2d}}{|\Lambda|} \sum_{\vec{x}\in\Lambda}
\int_{{\cal B}^2} \!\D \vec{q}\D\vec{q}'\: 
e^{\II (\vec{q}+\vec{q}')\cdot\vec{x}}\, \overline{\left\langle
\wit{S}(\vec{q},t)\wit{h}(\vec{q}')\right\rangle}
\nonumber \\
&=& \frac{(2\pi)^{-2d}}{|\Lambda|} \sum_{\vec{x}\in\Lambda}
\int_{{\cal B}^2} \!\D \vec{q}\D\vec{q}'\: 
e^{\II (\vec{q}+\vec{q}')\cdot\vec{x}} \int_{0}^t \!\D t'\:
e^{\omega(\vec{q})(t'-t)} \sqrt{\frac{g(t')}{g(t)}\,}\cdot
\overline{\wit{h}(\vec{q})\wit{h}(\vec{q}')}
\nonumber \\
&=& (2\pi)^{-d}\, 2\Gamma \int_{0}^t \!\D t'\: \int_{\cal B} \!\D\vec{q}\: 
e^{-\omega(\vec{q})(t-t')} \sqrt{\frac{g(t')}{g(t)}\,}
\nonumber \\
&=& {2\Gamma} \chi_{\rm FC}(t)
\EEA
as asserted, and where we used in the second line eq.~(\ref{3:gl:ssol}) and in 
the third line the field correlator (\ref{3:gl:varianzen}). 

After these preparations we can test our heuristic picture. Using the
techniques described in \cite{Pae03}, we obtain $g(t)$ by solving 
eq.~(\ref{3:gl:g}) numerically. 
In figure~\ref{Abb2} we show $\chi_{\rm FC}(t)$ for the three-dimensional
case, starting from an infinite-temperature initial state. We clearly see
that $\chi_{\rm FC}(t)$ saturates rapidly. Consequently, $A=0$ for all
temperatures $T\leq T_c$. Similar tests can be performed for other values
of $d$ as well. 

We pause a moment in order to discuss the functional form of the response
function. If $\Gamma=0$, solution of the spherical constraint gives for
$T<T_c$ the well-known exact result, valid for all values of $d$ and in
the ageing regime $s\gg 1$ and $t-s\gg 1$ is, 
see \cite{God00b,Zip00,Hen01,Pic02,Cor01,Cor03}
\BEQ \label{3:gl:Rex}
R(t,s) = r_0' \left(\frac{t}{s}\right)^{d/4} \left( t-s\right)^{-d/2}
\EEQ
with $r_0'=(4\pi)^{-d/2}$. From (\ref{1:gl:CRskal}), we read off $a=(d-2)/2$ and 
$\lambda_R/z=d/4$. We point out that this exact result for $a$ is in 
contradiction with the claim (\ref{1:gl:aHut}) raised by CLZ. Furthermore, we 
see that the exact result eq.~(\ref{3:gl:Rex}) has precisely the form 
(\ref{1:gl:lsi}) predicted by local scale invariance \cite{Hen01,Hen02}. 
A similar test can be performed for $T=T_c$ \cite{Hen01,Pic02}, or even in a 
spherical model with spatially long-ranged interactions \cite{Can01}. 
It is important to note that the local scale invariance 
prediction (\ref{1:gl:lsi}) applies to the full response
function $R(t,s)$ and not to the part remaining after subtraction of
a `stationary' term, as suggested in \cite{Cor03}.

{\bf 2.} As a second example, we give numerical evidence that $A=0$ in the $2D$ 
Glauber-Ising model quenched to its critical point $T=T_c$. We used a standard
heat-bath algorithm and measured the integrated linear response through 
eq.~(\ref{1:gl:Barrat}). Data for $\chi_{\rm FC}(t)$ thus obtained are 
displayed in figure~\ref{Abb4} and we see that saturation occurs. 

In summary, extending a heuristic argument of section~\ref{sect2}, we have
argued that for ferromagnetic systems in class L, one should find
saturation for the field-cooled susceptibility, 
viz. $\chi_{\rm FC}(t)\sim {\rm O}(1)$. We have confirmed this expectation 
in some models. 

The heuristic arguments in this and the preceeding section can only be applied
in $d\geq 2$ dimensions. Indeed, the ageing behaviour of the $1D$ Glauber-Ising 
model at its critical point $T_c=0$ is peculiar and will be 
discussed in the appendix.

\section{Scaling of the thermoremanent response} \label{sect4}

Since we have seen that the terms which describe the physically interesting 
ageing effects are only sub-leading in the zero-field-cooled susceptibility 
$\chi_{\rm ZFC}(t,s)$, we discuss in this section the scaling of the
thermoremanent susceptibility $\rho_{\rm TRM}(t,s)$. As we shall see, 
the case when the waiting time $s$ is small needs particular consideration.
It is well-known \cite{Hum91} that the response with respect to a fluctuation
in the initial state scales as
\BEQ
R(t,0) \sim t^{-\lambda_R/z}
\EEQ
On the other hand, it can be shown \cite{Zip00} that there is a time-scale
$t_p\sim s^{\zeta}$ with $0<\zeta<1$ such that if the time difference 
$\tau=t-s\lesssim t_p$, then the response function is still the one of the
equilibrium system $R(t,s)\simeq R_{\rm gg}(\tau)$ while scaling sets in 
and (\ref{1:gl:CRskal}) holds if $t\gtrsim t_p$. For example, in the spherical
model one has $\zeta=4/(d+2)$ \cite{Zip00}. We then have
\BEA
\rho_{\rm TRM}(t,s) &=& \int_0^s \!\D u\, R(t,u) = 
\int_0^s \!\D \tau\, R(t,s-\tau)
\nonumber \\
&=& \int_{0}^{t_p} \!\D\tau\, R_{\rm gg}(2\tau) + s \int_{t_p/s}^{t_{\eps}/s}
\!\D v\, R(t,s(1-v)) + \int_{t_{\eps}}^{s} \!\D \tau\, R(t,s-\tau)
\nonumber \\
&\simeq&  \int_{0}^{\infty} \!\D\tau\, R_{\rm gg}(2\tau)
+ s^{-a} \int_{0}^{1} \!\D v\, (1-v)^{-1-a} f_R\left(\frac{t/s}{1-v}\right)
+ c_{\infty}\, t^{-\lambda_R/z}
\EEA
Here, following \cite{Zip00,Cug02}, we have introduced a 
third time-scale $t_{\eps}$ such that 
$s-t_{\eps}={\rm O}(1)$. In the third line, the limit $s\to\infty$ is taken
and the last term is estimated from the mean-value theorem, where $c_{\infty}$
is a constant. Since we
are perturbing with a random magnetic field, the system is not driven to
a new equilibrium state and consequently, the first term
$\rho_{\infty} = \int_{0}^{\infty} \!\D\tau\, R_{\rm gg}(2\tau)$ vanishes if
the initial mean magnetization was taken to be zero. As a result we have
the scaling form, already announced in \cite{Hen02a}
\BEQ \label{4:gl:rhotrm}
\rho_{\rm TRM}(t,s) = s^{-a} f_{M}(t/s) + s^{-\lambda_R/z} g_M(t/s)
\EEQ
where $f_M(x)$ and $g_M(x)$ are scaling functions. These results only depend
on the assumption of dynamical scaling. If in addition local scale invariance
applies, the form of the response function $R(t,s)$ is given by 
eq.~(\ref{1:gl:lsi}) and therefore \cite{Hen02,Hen02a}
\BEQ
f_M(x) = r_0\, x^{-\lambda_R/z} {_2F_1}\left(1+a,\frac{\lambda_R}{z}-a;
\frac{\lambda_R}{z}-a+1; \frac{1}{x}\right) \;\; , \;\;
g_M(x) \simeq r_1 x^{-\lambda_R/z}
\EEQ
and where $r_{0,1}$ are non-universal constants and ${_2F_1}$ is a 
hypergeometric function \cite{Notiz1}. 
 
The importance of taking this finite-time correction into account is 
illustrated in figure~\ref{Abb5}. For a fixed value of $x=t/s$ and $T<T_c$, 
we plot data for the Glauber-Ising model in $2D$ and $3D$, respectively and 
the exact solution of the Langevin equation of the spherical model, 
with $\Gamma=0$. A fully disordered initial state was used. 
We compare the data with the leading scaling form 
$\rho(t,s)\sim s^{-a}$ which for the times accessible does not fully describe
the data, but inclusion of the second term in (\ref{4:gl:rhotrm}) gives a
very good fit. From this fit we find the non-universal values of
$r_{0,1}$ listed in table~\ref{tab2}. In order to achieve this, however,
$a$ must take the values given in table~\ref{tab1}. 
Specifically, for
the Glauber-Ising model with $d\geq 2$ (which is in class S), 
we must take $a=1/2$, while if we had chosen $a=1/4$ as advocated by CLZ, 
only a fit of very low quality is obtained, see \cite{Hen02a,Hen03a}. 
This provides further evidence against the proposed eq.~(\ref{1:gl:aHut}).
We point out that having fixed $r_{0,1}$
for a given value of $x$, the scaling of $\rho_{\rm TRM}(t,s)$ is completely
fixed and does not contain any further tunable parameter. Repeating the 
comparison between the data and (\ref{4:gl:rhotrm}) for other values of $x$
therefore allows to test the predictions of local scale invariance and this
has been carried out in detail for the $2D$ and $3D$ Glauber-Ising model
\cite{Hen02a,Hen03b}. While in the $2D$ Glauber-Ising model the curvature of 
the data, see figure~\ref{Abb5}a, already suggests
the presence of strong finite-time corrections to scaling, in $3D$ it might
appear at first sight that a reliable value of the
exponent $a$ could be derived, see figure~\ref{Abb5}b. However, the inclusion
of the leading correction term from eq.~(\ref{4:gl:rhotrm}) 
together with the study of the $x$-dependence of the coefficients, as carried 
out in \cite{Hen02a}, is required in order to avoid a systematic error in the 
determination of $a$. In this way the expected value $a=1/z=1/2$ is 
reconfirmed. 
Carrying out the analysis of the scaling functions as described in
\cite{Hen02a}, we can conclude that local scale invariance holds
in the $2D$ and $3D$ Glauber-Ising model with $T<T_c$. 

A closer inspection of table~\ref{tab2} shows that in the spherical model with
$d>4$ and $T<T_c$, the signs of $r_0$ and $r_1$ change and that the term which 
was treated as a correction in (\ref{4:gl:rhotrm}) becomes dominant. Indeed, 
the exactly known exponents $a=(d-2)/2$ and $\lambda_R/z=d/4$ 
cross at the upper critical dimension $d^*=4$. We conclude that the scaling 
argument used in \cite{Hen02a} to derive the relation $a=(d-2+\eta)/z$ for 
class L systems should rather be viewed as a hyperscaling argument. 

\section{Conclusions}

In this paper, we have analysed the long-time behaviour of integrated
response functions in ageing spin systems without disorder. While from the
scaling form (\ref{1:gl:CRskal}), one might have expected a simple scaling
$\chi(t,s)\sim s^{-a}$ of the integrated response, we have shown that matters
are more complicated. Conceptually, the issue can be clarified by studying
the scaling of the {\em field-cooled} susceptibility $\chi_{\rm FC}(t)$ and
we have seen that two broad classes of systems must be distinguished, which
are called classes S and L, according to whether their equilibrium spin-spin
correlator shows short-ranged or long-ranged spatial decay, respectively
\cite{Hen02a}. Specifically, we have found
\begin{enumerate}
\item 
For systems of class S, we have
\BEQ \label{5:gl:chiFC}
\chi_{\rm FC}(t) -\chi_0 \sim t^{-A}
\EEQ
where $T \chi_0=1-m_{\rm eq}^2$ and $m_{\rm eq}$ is the equilibrium 
magnetization.The exponent $A$ is related to the interface width exponent 
$\kappa$. If $w(t)\sim t^{\kappa}$, we have seen that
\BEQ
A = \frac{1}{z} - \kappa \;\; ; \;\;
\mbox{\rm ~~(class S)}
\EEQ
but provided that the temperature $T>T_R$ is above the roughening temperature
$T_R$. 
\item For systems of class L, eq.~(\ref{5:gl:chiFC}) still holds with
$\chi_0=0$, but with
\BEQ
A = 0 \;\; ; \;\;
\mbox{\rm ~~(class L)}
\EEQ
\item 
The scaling of the thermoremanent susceptibility $\rho_{\rm TRM}(t,s)$
may be affected by a relatively large finite-time correction term, see
eq.~(\ref{4:gl:rhotrm}) where the associated scaling functions can be found
explicitly from local scale invariance. 
\item 
Consequently, the leading time-dependence of the field-cooled susceptiblity 
$\chi_{\rm ZFC}(t,s)-\chi_0\sim \chi_{\rm FC}(t)-\chi_0\sim t^{-A}$, whereas
the ageing terms expected from integrating eq.~(\ref{1:gl:CRskal}) merely arise
as finite-time corrections.
\end{enumerate}
Our results reproduce the entirety of known results in ageing ferromagnetic
spin systems with $d \geq 2$ but do not require to postulate an upper 
critical dimension $d=3$ in the Glauber-Ising model. For systems of class S, 
the exponent relation $a=1/z$ is reconfirmed while the proposed 
eq.~(\ref{1:gl:aHut}) is invalidated. 

If questions of simulational efficiency play no r\^ole, 
it might be technically easier to avoid both the ZFC and the TRM protocol, 
as already suggested in \cite{Cor03}. For example,
we propose the `intermediate' protocol, which runs as follows: quench the
system at $t=0$ without a magnetic field and fix a waiting time $s$. 
At time $s/2$, turn on a random magnetic field and keep in on until the
waiting time $s$. Then turn the field off again and measure the magnetization
at the observation time $t>s$. The {\it intermediate integrated response} is
\BEQ
\rho_{\rm INT}(t,s) = M_{\rm INT}(t,s)/h := 
\int_{s/2}^{s} \!\D u\, R(t,u) = s^{-a} f_{\rm INT}(t/s)
\left( 1 + {\rm o}(s^{-\lambda_R/z})\right)
\EEQ
and should be free of the leading term coming from the interface roughness
as well as the finite-time correction of order ${\rm O}(s^{-\lambda_R/z})$. 
We illustrate this for the mean spherical model in figure~\ref{Abb6}, where it 
can be seen that already for times much shorter than in figure~\ref{Abb5}c, the
{\em linear} response (\ref{3:gl:chifc}) obtained from the exact solution 
of the Langevin equation (with $\Gamma=0$)
converges to the expected power law, with $a=0.5$ in $3D$. 
It would be interesting to see if a recent method to calculate 
$\chi_{\rm ZFC}(t,s)$ directly in the Glauber-Ising model \cite{Ric03} 
(which in turn is based on a method to estimate $R(t,s)$ in an Ising model
with a different dynamics modified from Glauber dynamics \cite{Cha03}) 
could be generalized to find $\rho_{\rm INT}(t,s)$ as well. 

\zeile{2}
\noindent {\bf Acknowledgements:} 
It is a pleasure to thank G.T. Barkema for a very useful discussion,
C. Godr\`eche, J.-M. Luck and G.M. Sch\"utz for helpful conversations and the 
Max-Planck Institut f\"ur Physik komplexer Systeme (MPIPKS) in Dresden for warm 
hospitality during the NESPHY03 Seminar, where this work was done. 
This work was supported by the Bayerisch-Franz\"osisches Hochschulzentrum 
(BFHZ), by CINES Montpellier (projet pmn2095), and by the Regionales
Rechenzentrum Erlangen (through the use of the IA32 Compute-Cluster).

\appsection{}

In the main text, we studied the time-dependent scaling of the
integrated response $\chi_{\rm FC}(t)$. We shall now attempt to extract 
further information on the scaling of $R(t,s)$ when $t\approx s$. 
Note that the scaling form eq.~(\ref{1:gl:CRskal}) 
{\em cannot} be used close to the upper integration limit, since 
there the condition $\tau=t-s\gg 1$ needed for the validity of 
(\ref{1:gl:CRskal}) does not hold. Rather, one might argue that there should be 
some time-scale $t^*$ such that (\ref{1:gl:CRskal}) holds for waiting times 
$s\lesssim t^*$. If in addition $t-t^*$ is some `small' constant $\Delta t^*$ 
for large times, we may write 
\BEA
\chi_{\rm FC}(t) &=& \int_0^{t^*}\!\D s\, R(t,s) + \int_{t^*}^t\!\D s\, R(t,s)
\nonumber \\
&\simeq& t^{-a} \int_{t/t^*}^{\infty}\!\D v\, v^{a-1} f_R(v) +
(t-t^*) R(t,t')
\nonumber \\
&\simeq& t^{-a} \int_1^{\infty} \!\D v\, v^{a-1} f_R(v) 
+ \Delta t^* \cdot R(t,t')
\label{A:Rtt}
\EEA
where we used the mean-value theorem and $t'\in [t^*,t]$. If
the second term in (\ref{A:Rtt}) is the leading one, then comparison with 
the scaling of $\chi_{\rm FC}$ gives 
$R(t,t)\simeq 1 + {\rm O}(t^{-A})$, provided the 
conditions stated above are satisfied. Here the value of the constant is
fixed from the physical consideration that for $\tau =t-s\ll 1$ the system 
should still be in quasiequilibrium. 

We can now try and see to what extent this argument applies in specific models 
and to what extent it might be justified. The only model of class S we studied 
here is the $2D$ Glauber-Ising model at $T<T_c$. Then $a=1/2$ and $A=1/4$ and 
the contributions close to the upper integration limit are indeed dominant. One
should therefore expect 
$\lim_{\eps\to 0} R(t,t-\eps)\simeq {\rm cste.} + {\rm O}(t^{-1/4})$ for
$t$ large. 

On the other hand, we have considered in section~\ref{sect3} several examples 
of systems of class L, where we have found $A=0$ throughout. 
First, for the spherical model, the exact expression for $R(t,s)$ gives
$R(t,t-2\eps)=f(\eps)\sqrt{g(t-2\eps)/g(t)}\simeq 
1+\eps \left(f'(0)-g'(t)/g(t)\right)+{\rm O}(\eps^2)$. Since for
$t$ large, one has $g(t)\sim t^{\psi}$ with a known exponent $\psi$ 
\cite{God00b,Pic02}, we obtain for $\eps\to 0$
$R(t,t-2\eps)\simeq {\rm cste.} + {\rm O}(t^{-1})$, and have indeed $A=0$,
since the constant is different from unity. 
Second, a similar conclusion should also hold true for the $2D$ Glauber-Ising 
model at criticality. 
Third, an analogous result can also
be obtained in the $2D$ linear voter model at criticality \cite{Sas03}, where
the exact response function $\lim_{\eps\to 0}R(t+\eps,t)\sim (\ln t)^{-1}$. 

Additional studies to test these conclusions further would be welcome.

The heuristic arguments presented in this paper assume $d\geq 2$ space 
dimensions throughout. As an illustration what can happen in one dimension,
we now consider the $1D$ Glauber-Ising model at its critical 
temperature $T_c=0$. This model has been investigated
countless times and we shall not repeat the details
of its definition (see e.g. \cite{God00a,Hen03d} and references therein) here.
The exact autoresponse function is
\BEQ \label{A:gl:Rts}
R(t,s) = e^{-(t-s)} I_0(t-s) \left( 1 - C_1(s)\right)
\EEQ 
where $C_1(s)$ is the equal-time correlator of two spins on neighbouring
sites. For an infinite-temperature initial state, one has \cite{Hen03d}
\BEQ \label{3:gl:C1}
C_1(s) = e^{-2s}\left[ I_1(2s) + 2\sum_{n=1}^{\infty} I_{n+1}(2s) \right]
= 1- e^{-2s}\left[ I_0(2s) + I_1(2s)\right]
\EEQ
where in the second step the Bessel-function identity
$e^z = I_0(z) + 2\sum_{n=1}^{\infty} I_n(z)$ was used. It has recently been 
shown that for any initial correlators of the power-law form 
$C_{\rm ini}(r)\sim r^{-\nu}$ with $\nu\geq 0$ the leading scaling behaviour 
of $C_1(s)$ for $s$ large is unchanged \cite{Hen03d}. 
Therefore, for any power-law correlated initial state we have for $t$ large 
\BEQ \label{A:gl:R}
\lim_{\eps\to 0}R(t,t-\eps)\simeq \frac{1}{\sqrt{\pi}}\cdot t^{-1/2}
\EEQ
Comparison of eqs.~(\ref{A:Rtt}) and (\ref{A:gl:R}), if admissible, 
would then give $A=1/2$. 
Such a result would appear to be natural in the setting of section~\ref{sect2} 
if we recall that at $T=0$ there are fully ordered domains
where the width of the domain walls is one lattice constant, 
thus $w(t)={\rm cste.}$

On the other hand, integrating (\ref{A:gl:Rts},\ref{3:gl:C1}), the field-cooled 
susceptibility becomes for an uncorrelated initial state
\BEQ \label{3:gl:GI1d}
\chi_{\rm FC}(t) = \int_0^t\!\D s\, e^{-t-s} I_0(t-s) \left(
I_0(2s)+I_1(2s)\right) 
\EEQ
and we show this in figure~\ref{Abb3}. Clearly, there is 
saturation for large times and in the spirit of section~\ref{sect3} one would
conclude $A=0$. Therefore, a na\"{\i}ve application of the arguments valid
for $d\geq 2$ would lead to two different values of $A$ in this $1D$ model. 
We leave open the question how this apparent inconsistency might be resolved.


\newpage 

\begin{table}[h]
\caption{Non-equilibrium exponents $a$ and $b$ for quenches from a fully
disordered state onto and below the critical point $T_c>0$ of simple 
ferromagnets of the classes S and L defined in the text, 
according to {\protect\cite{Hen02a}}. \label{tab1}}
\begin{tabular}{|c|cc|c|}  \hline
        & $a$            & $b$            & Class \\ \hline
$T=T_c$ & $(d-2+\eta)/z$ & $(d-2+\eta)/z$ & L \\ \hline
$T<T_c$ & $(d-2+\eta)/z$ & 0              & L \\
        & $1/z$          & 0              & S \\ \hline
\end{tabular}
\end{table}

\begin{table}[h]
\caption{Values of the autoresponse exponent $\lambda_R$ 
and of the parameters $r_0$ and $r_1$ in the 
Glauber-Ising model in two dimensions at $T=1.5$ and in 
three dimensions at $T=3$ and of the mean spherical model at $T=2$ in several
dimensions. An infinite-temperature initial state was used. The
dynamic exponent $z=2$. \label{tab2}}
\begin{tabular}{|l|c|cll|}  \hline
model     & $d$ & $\lambda_R$ & \multicolumn{1}{c}{$r_0$} & 
\multicolumn{1}{c|}{$r_1$}                   \\\hline
Glauber-Ising & 2   & 1.26 & $~~~1.76 \pm 0.03$ & $-1.84 \pm 0.03$  \\ 
          & 3   & 1.60  & $~~~0.10 \pm 0.01$  & $~~\,0.20\pm 0.01$ \\\hline
spherical & 3   & 1.50  & $~~\:0.180\pm 0.01$ & $-0.081\pm 0.002$  \\
          & 3.5 & 1.75  & $~~\:0.20~ \pm 0.01$ & $-0.126\pm 0.003$  \\
          & 4.5 & 2.25  & $-0.056\,\pm 0.003$   & $~~\:0.095\pm 0.002$ \\\hline
\end{tabular}
\end{table}

\newpage 

{
\begin{figure}[htb]
\centerline{\epsfxsize=4.25in\ \epsfbox{
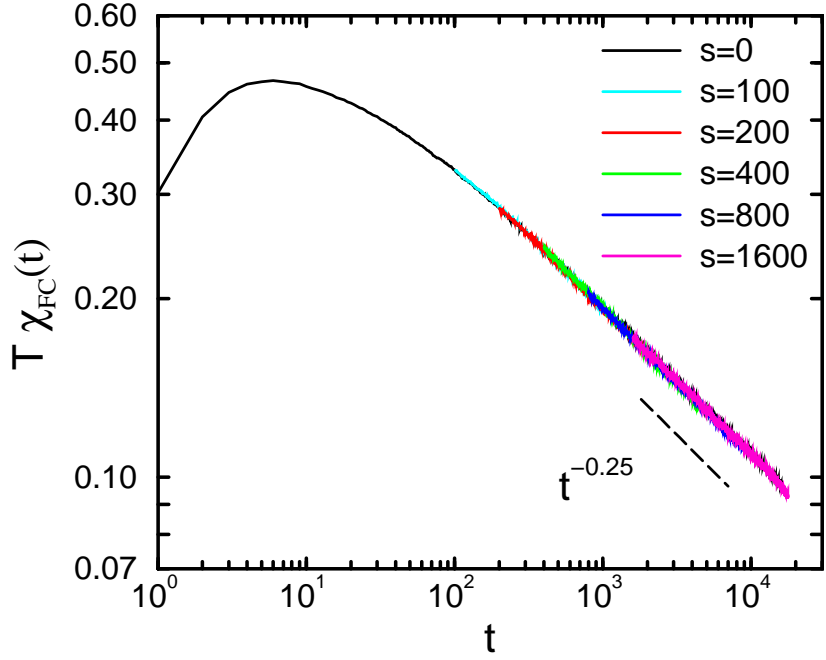} }
\caption{Field-cooled susceptibility $\chi_{\rm FC}(t)=
\rho_{\rm TRM}(t,s)+\chi_{\rm ZFC}(t,s)$ in the $2D$ Glauber-Ising model
at temperature $T=1.5<T_c$ and for several waiting times. An alternating 
field $h_{\vec{i}}=\pm h$ with $h=0.05$ was used. The power law
$\chi_{\rm FC}(t)\sim t^{-0.25}$ is also shown for comparison. 
The simulated systems contained $300 \times 300$ spins, and the data 
have been obtained after averaging over at least 5000 different runs.
\label{Abb1}}
\end{figure}
}
\zeile{2}
{
\begin{figure}[htb]
\centerline{\epsfxsize=3.85in\ \epsfbox{
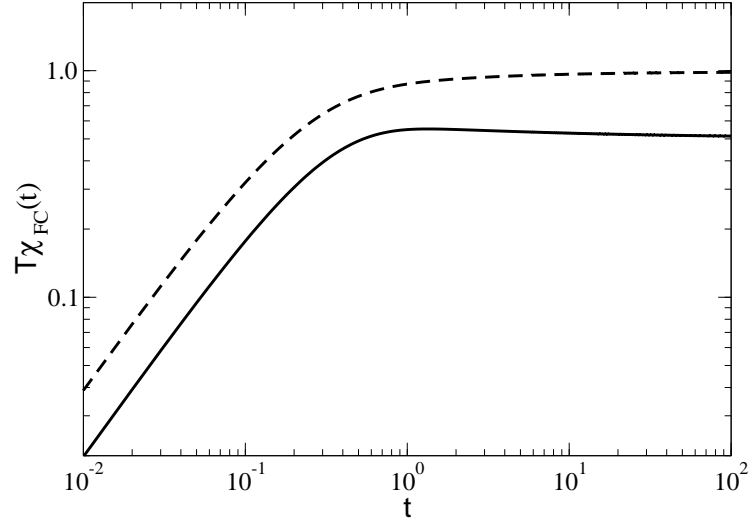} }
\caption{~\\Field-cooled susceptibility $\chi_{\rm FC}(t)$ for the
$3D$ kinetic mean spherical model in a gaussian random magnetic field of width
$\Gamma=0.01$ and at $T=2<T_c$ (full curve) and at $T=T_c\simeq3.96$ 
(dashed curve). 
\label{Abb2}}
\end{figure}
}
\zeile{2}
{
\begin{figure}[htb]
\centerline{\epsfxsize=4.25in\ \epsfbox{
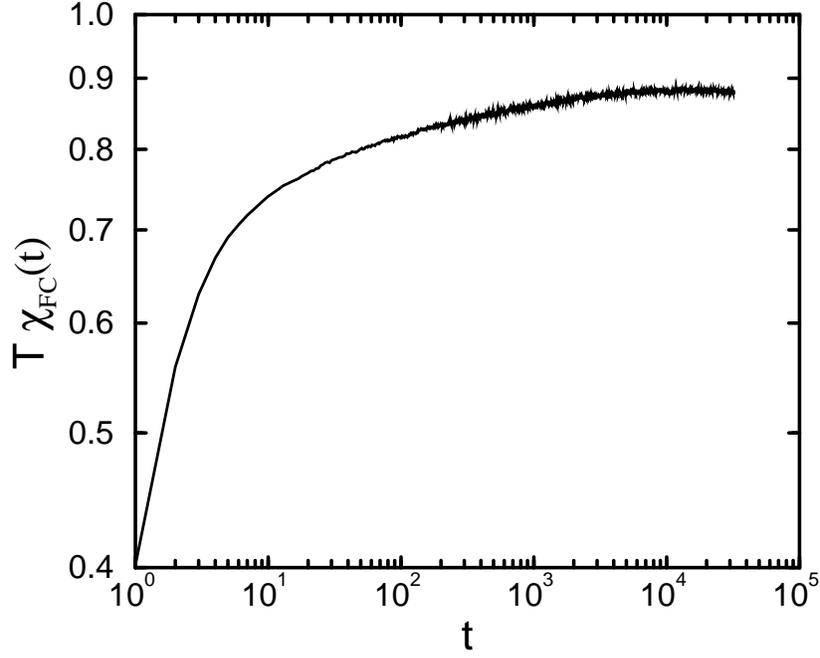} }
\caption{Field-cooled susceptibility $\chi_{\rm FC}(t)$ in the $2D$ 
Glauber-Ising model at the critical temperature $T=T_c\approx 2.269$. 
An alternating field $h_{\vec{i}}=\pm h$ with $h=0.05$ was used.
Systems with $300 \times 300$ spins have been simulated.
\label{Abb4}}
\end{figure}
}
\zeile{2}
{
\begin{figure}[htb]
\centerline{\epsfxsize=4.0in\ \epsfbox{
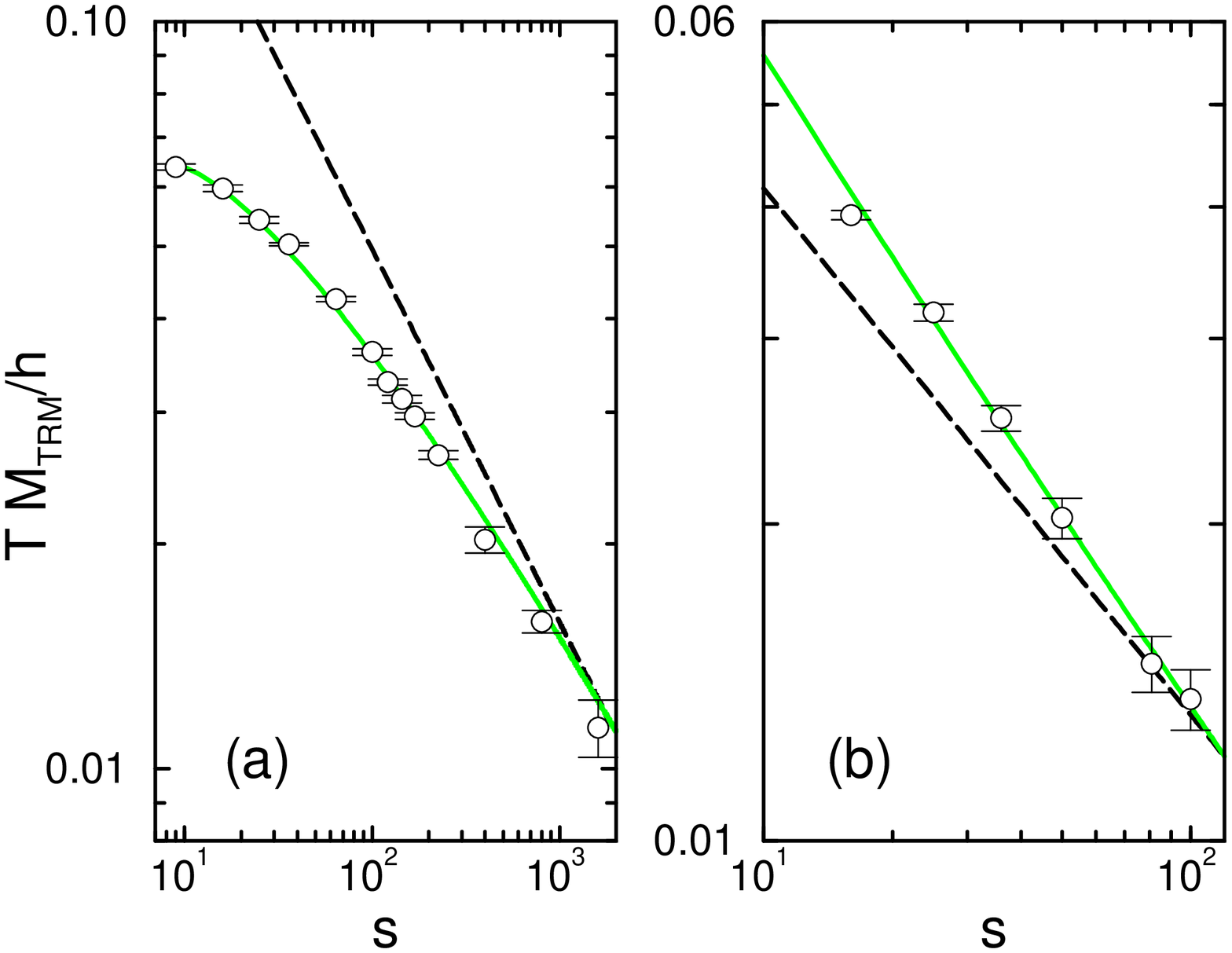} ~~
\epsfxsize=1.7in\epsfbox{
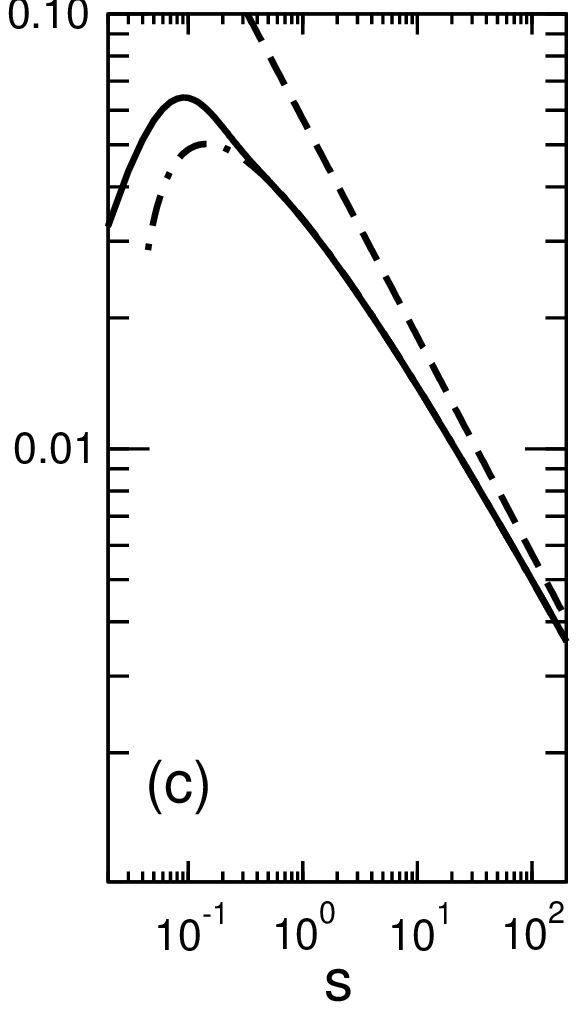}
}
\caption[Scaling of the TRM]{Scaling of the thermoremanent susceptibility 
$\rho(t,s)=M_{\rm TRM}(t,s)/h$ as a function of the waiting time $s$ for a 
fixed value of $x=t/s$ for three models. (a) The $2D$ Glauber-Ising
model at $T=1.5$ and with $x=7$ (points) is compared with the scaling 
prediction eq.~(\ref{4:gl:rhotrm}) (full grey curve) and with the simple 
scaling form $\rho(t,s)\sim s^{-1/2}$ (dashed curve). 
(b) Same comparison for the $3D$ Glauber-Ising model at $T=3$ and with $x=5$. 
(c) The $3D$ spherical model at $T=2$ and $x=5$. The full curve is the 
exact solution of the Langevin equation, the dash-dotted line the scaling
prediction eq.~(\ref{4:gl:rhotrm}) and the dashed line the simple form
$\rho(t,s)\sim s^{-1/2}$.
\label{Abb5}}
\end{figure}
}
\zeile{2}
{
\begin{figure}[htb]
\centerline{\epsfxsize=4.25in\ \epsfbox{
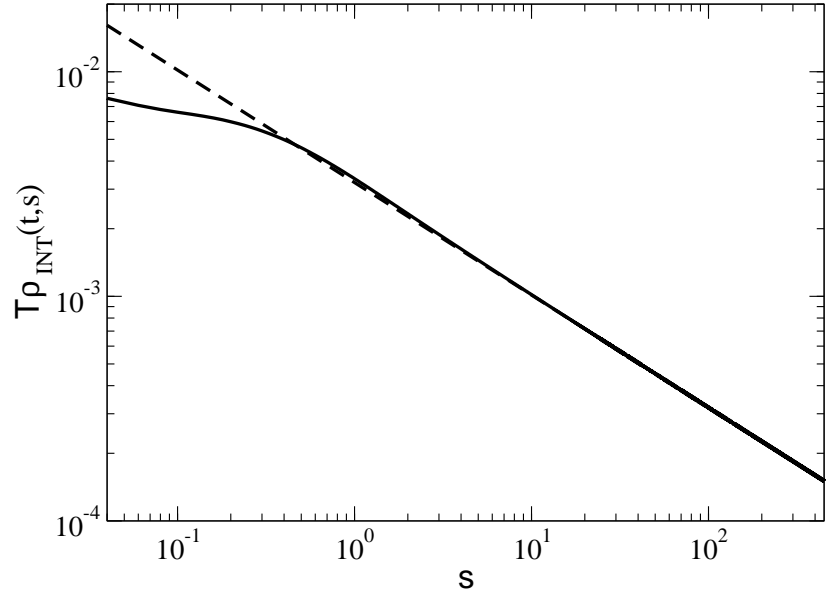} }
\caption{Integrated linear reponse $\rho_{\rm INT}(t,s)$ according to the 
intermediate protocol for the $3D$ mean spherical model with $T=2$, 
an infinite-temperature initial state and $x=20$. The full curve is the
exact solution, the dashed line is a power-law fit $\sim s^{-0.5}$.   
\label{Abb6}}
\end{figure}
}
\zeile{2}
{
\begin{figure}[htb]
\centerline{\epsfxsize=4.25in\ \epsfbox{
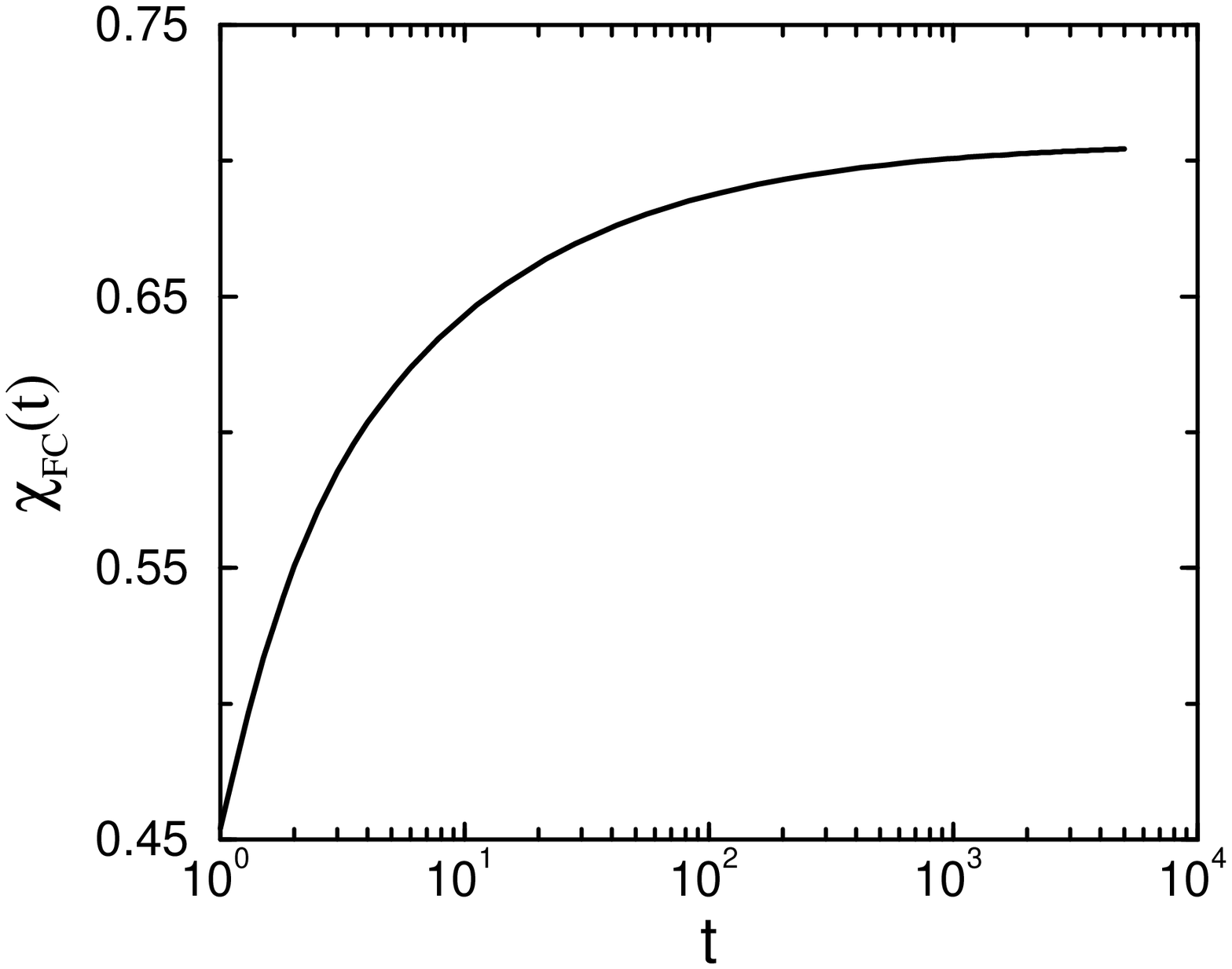} }
\caption{Field-cooled susceptibility $\chi_{\rm FC}(t)$ for the
$1D$ Glauber-Ising model at $T=0$. 
\label{Abb3}}
\end{figure}
}


\begin{thebibliography}{999}
\bibitem{Notiz1} CLZ also attempt to calculate $f_{M}(x)$, but 
in their calculation they take the large-$x$ limit 
$f_{R}(x)\sim x^{-\lambda_R/z}$ before integrating and approximate the 
small-$x$ behaviour of $f_M(x)$ by a constant \cite[eq.(83)]{Cor03}. That is
too rough an approximation to be useful.
\end{thebibliography}
\end{document}